# Quantifying Global Food Trade: A Net Caloric Content Approach to Food Trade Network Analysis

**Enter authors here: Xiaopeng Wang[1], Chengyi Tu[2], Shuhao Chen[2], Sicheng Wang[2], Ying Fan[3], Samir Suweis[4] and Paolo D'Odorico[5]**

[1] Modern Business Research Center, School of Management and E-Business, Zhejiang Gongshang University, 310018 Hangzhou, China.

[2] School of Economics and Management, Zhejiang Sci-Tech University, Hangzhou 310018, China.

[3] College of Geography and Environment, Shandong Normal University, Jinan 250358, China.

[4] Department of Physics and Astronomy "G. Galilei", University of Padua, 35131 Padova, Italy.

[5] Department of Environmental Science, Policy, and Management, University of California, Berkeley, CA 94720-3114, USA.

Corresponding author: Chengyi Tu (chengyitu1986@gmail.com) and Paolo D'Odorico (paolododo@berkeley.edu)

**Key Points:**

- Global food trade analysis highlights calorie content, offering insights beyond trade volume for food security and policy discussions.

- Increased network connectivity suggests improved resilience, enabling more efficient food redistribution during shortages.

- Trade network modularity has risen, reflecting growing regional trade clusters, which may influence global food system stability.

**Abstract**

As the global population and the per capita demand for resource intensive diets continues to grow, the corresponding increase in food demand challenges the global food system, enhancing its reliance on trade. Most previous research typically constructed either unweighted networks or weighted solely by tonnage to represent food trade, and focused on bilateral trade relationships between pairs of countries. This study investigates the properties of global food trade constructed in terms of total food calories associated with all the main food products exchanged along each trade link (edge of the food trade network). Utilizing data from the Food and Agriculture Organization between 1986 and 2022, we construct a directed, weighted network of net caloric flows between countries. This approach highlights the importance of considering nutritional value in discussions of food security and trade policies, offering a more holistic view of global food trade dynamics. Our analysis reveals significant heterogeneity in trade patterns, with certain countries emerging as major exporters or importers of food calories. Moreover, we employ network measures, including network connectivity, network heterogeneity, network modularity, and node correlation similarity, to elucidate the structural dynamics of global net food calorie trade networks that are relevant to the stability and resilience of the global food system. Our work provides a more nuanced understanding of global food trade dynamics, emphasizing the need for comprehensive strategies to enhance the resilience and sustainability of food trade networks.

**Plain Language Summary**

As global food demand rises, understanding how food trade networks distribute calories is essential for ensuring global food security. This study analyzes the global food trade from 1986 to 2022 by focusing on the net calorie content, rather than just the mass, of traded food items. By converting trade data into calorie units, the authors built a global network map that reveals which countries are primary exporters and importers of food calories. This approach offers a fresh perspective on the resilience of food trade networks, emphasizing the nutritional value exchanged across countries. The analysis shows that some countries, like the United States and Brazil, are major calorie exporters, while others, like China and Japan, are significant importers.

The study also highlights "peripheral" countries that rely heavily on imports to meet their nutritional needs. Over time, the network has become more connected and clustered, pointing to a trend of regionalized trade. This structure provides insights into the stability of food supply systems, indicating potential risks for food security due to trade disruptions. The findings underscore the importance of considering both the quantity and nutritional quality of traded food in addressing global food security challenges.

**1 Introduction**

As the global population continues to grow and emerging economies are adopting more resource-intensive diets richer in meat and other animal products (Boakes et al., 2024; D'Odorico et al., 2018; Dalin & Outhwaite, 2019; Springmann et al., 2023), the corresponding increase in food demand challenges the global food system and its reliance on trade (Alexandratos & Bruinsma, 2012; Falcon et al., 2022; Giller et al., 2021; Oluwole et al., 2023). This anticipated surge necessitates a comprehensive and multifaceted approach to effectively address the resultant environmental, economic, and social impacts (Read et al., 2020; Stein & Santini, 2022; Vågsholm et al., 2020; Z. Xu et al., 2020). Ensuring access to sufficient and nutritious food is not only a fundamental human right but also a cornerstone of the United Nations' Sustainable Development Goals (Carlsen & Bruggemann, 2022; Jia et al., 2022; Nakai, 2018). The current structure of global food trade plays a pivotal role in facilitating nutrient access, particularly for poorer nations, enabling them to nourish hundreds of millions of people (Ardan et al., 2023; Deniz Berfin Karakoc et al., 2023; Deniz Berfin Karakoc et al., 2022; Konar et al., 2018; Laborde et al., 2020; Silvestrini et al., 2023). However, this system also brings forth complex issues such as the disconnection between producers and consumers and the loss of resilience associated with many countries depending on resources they do not control (D'Odorico et al., 2019; Dalin et al., 2012; Konar et al., 2011; Konar et al., 2018; Mekonnen et al., 2024; Suweis et al., 2015; Suweis et al., 2013).

Complex networks are essential for modeling global food trade, linking countries around the world (Carr et al., 2012; D'Odorico et al., 2014; D'Odorico et al., 2012; Ercsey-Ravasz et al., 2012; Suweis et al., 2011). At the core of the global food trade network lies a dynamic interplay of economic, environmental, and social factors. The network's structure is characterized by a dense web of bilateral trade relationships, where countries act as nodes connected by trade flows. These connections are influenced by a several factors, including geopolitical relations, trade agreements, countries' availability of suitable agricultural land, water availability, population size, dietary trends, economic development, and comparative advantages in agricultural production (Barigozzi et al., 2011; Suweis et al., 2011). The resilience and efficiency of these networks are critical in mitigating the impacts of local production shocks and ensuring a stable supply of food across the globe (Chengyi Tu et al., 2019). Recent studies have underscored the dual role of global food trade networks in both mitigating and propagating risks (Chaudhuri et al., 2016; Gomez et al., 2021). On the one hand, international trade allows countries to buffer against local production shortfalls by importing food from other regions, thus enhancing food security (Gomez et al., 2021; Halpern et al., 2022; Hicks et al., 2022; Suweis et al., 2013). On the other hand, the interconnected nature of these networks can also facilitate the rapid spread of disruptions, such as those caused by climate change, economic crises, or geopolitical conflicts (Chung & Liu, 2022; Kreibich et al., 2022; Marchand et al., 2016; Mehrabi et al., 2022; Tamea et al., 2014). Understanding the topological structure and dynamics of these networks is therefore essential for developing strategies to enhance their resilience and sustainability (Davis et al., 2021; Muller et al., 2021; Chengyi Tu et al., 2019). Recent work on the effect of network topology on the stability and resilience of global food security has shown how an increase in network connectivity (i.e., globalization) or modularity (i.e., regionalization) may either increase or decrease the resilience, depending on whether the network is random or heterogeneous (i.e., with a negative correlation between the number incoming or outgoing connections at each node) (Chengyi Tu et al., 2019).

Despite the extensive body of research on global food trade networks, a significant limitation persists: the predominant reliance on either unweighted networks or on networks weighted solely by tonnage (D. B. Karakoc & Konar, 2021; Oluwole et al., 2023; Silvestrini et al., 2023; H. Xu et al., 2024; Z. Xu et al., 2020). While this approach offers valuable insights, it fails to capture the full complexity and diversity of food items traded internationally (C. Tu et al., 2016). Given the vast array of food products, each with distinct economic, nutritional, and environmental implications, it is imperative to establish a standardized metric for quantification. Such a metric would enable a more nuanced and comprehensive analysis of food trade flows between countries. Moreover, most research tends to focus on the bilateral trade relationships between pairs of countries (Oluwole et al., 2023; Silvestrini et al., 2023; H. Xu et al., 2024), representing each country as a node in the network, with connections directed by the flow of food trade. This method primarily accounts for the volume of trade, neglecting the critical aspect of net nutritional energy.

In this paper, following our previous work (D'Odorico et al., 2014; Chengyi Tu et al., 2019), we express global food trade in terms of caloric content. This approach allows for an analysis of global food trade dynamics that is more directly relevant to the food security. Indeed, by constructing a directed, weighted network of net caloric flows between countries, we provide new insights into recent changes in food security dependence on global food trade. Our findings reveal significant heterogeneity in trade patterns, with certain countries emerging as major exporters or importers of calories. Moreover, we employ network measures, including network connectivity, network heterogeneity, network modularity, and node correlation similarity, to elucidate the structural dynamics of global food trade networks. The stability of these patterns over time underscores the enduring influence of geographical and economic factors on global food trade. This study highlights the importance of considering nutritional value in discussions of food security and trade policies, offering a more holistic view of global food trade dynamics.

**2 Materials and Methods**

2.1 Unifying traded food to calorie content

In our study, we compile a comprehensive dataset of global food trade spanning from 1986 to 2022, sourced from the Statistics Division of the Food and Agriculture Organization (FAOSTAT) as of July 2024 (Organization, 2024b). To mitigate the impact of statistical discrepancies and data asymmetries, we have exclusively utilized export data reported by each country. The record format is exemplified as follows: country i exports 1 tons of bran wheat to country j.

Given the vast array of food items traded globally, it is essential to establish a consistent metric for quantification to measure food trade between countries. To this end, we propose converting the weight of all food items into a unified measure of caloric content. This conversion is executed using the Nutritive Factors (Organization, 2024a), a resource published by the Food and Agriculture Organization's Economic and Social Development stream. This tool enables us to convert each food item from its original unit (tons) into kilocalories, thereby providing a standardized measure of nutritional energy across the spectrum of food items. The record format post-conversion is as follows: country i exports 1 tons of bran wheat to country j, which, according to the Nutritive Factors, equates to 3300 kilocalories (given that 100 grams of bran wheat contains 330 calories), defined by $c_{ji}^{branwheat} = 3300$. Due to the substantial volume of trade involving feed crops (i.e., crops consumed by livestock rather than directly by humans), all secondary products are excluded from the calculation of caloric content. However, an exception is made for animal-based food products such as meat, milk, and eggs, which are classified as primary products.

Upon the completion of this conversion process, each instance of a food item being exported from country i to country j is documented as an individual record. These records are subsequently consolidated to compute the aggregate caloric value of the exports flowing from country i to country j. The record format is as follows: $c_{ji} = \sum_{p} c_{ji}^{p}$, where p includes all available food items. This approach extends beyond the conventional focus on trade volume, integrating the nutritional value of the traded goods into the analysis.

2.2 Constructing the weighted food trade network

The global food trade network is characterized by both weighted and directed attributes. Within this complex network, each country participating in global food trade is represented by a node, with connections between nodes directed by the flow of food trade. We focus on the net caloric flow between pairs of countries, which provides valuable insights into a country's food security status. For instance, a country that is a net importer of food calories may be more susceptible to food supply shocks from other countries. The net caloric flow can also highlight potential trade imbalances, for example, when some countries export high-calorie foods while importing low-calorie foods. Additionally, it is possible for two countries to trade similar volumes of food, yet the caloric content can vary significantly. By focusing on the net caloric exchange, we can gain a more comprehensive understanding of global food trade dynamics.

Specifically, we quantify the net caloric flow between countries using the metric $c_{ij} - c_{ji}$, where $c_{ij}$ and $c_{ji}$ denote the caloric flow from country j to country i and from country i to country j, respectively. A positive value of this metric indicates that country j exports more calories to country i than it imports from it. Consequently, an edge is established from country j to country i within the network, with a weight of $w_{ij} = c_{ij} - c_{ji}$, representing the net caloric flow, and $a_{ji} = 0$, indicating the absence of an edge from country i to country j. Conversely, a negative value signifies that country i exports more calories to country j than it imports from it. In this scenario, an edge is formed from country i to country j, with a weight of $w_{ji} = |c_{ij} - c_{ji}|$, again representing the net caloric flow, and $w_{ij} = 0$, indicating the absence of an edge from country j to country i. This methodological approach enables the construction of a directed, weighted network that accurately encapsulates the net caloric flows between countries, thereby providing profound insights into the dynamics of global food trade.

2.3 Summary of construction steps

In brief, the construction of the global net food calorie trade networks can be summarized through the following steps (see Fig. 1):

Step 1 Conversion to Caloric Content: Each export record for different food items is converted from its original weight into a unified measure of caloric content, denoted as $c_{ji}^{p}$ where p represents a food item, i and j are export and import country.

Step 2 Aggregation of Export Records: All caloric export record from country i to j are aggregated to calculate the total export caloric from country i to j, denoted as $c_{ji} = \sum_{p} c_{ji}^{p}$.

Step 3 Calculation of Net Caloric Export: The net caloric export from country j to i is calculated as $w_{ij} = c_{ij} - c_{ji}$ if the result is positive and its corresponding unweighted edge is $a_{ij} = 1$; in that case we set $w_{ji} = 0$ and $a_{ji} = 0$. Conversely, a net caloric import ($w_{ji} = |c_{ij} - c_{ji}|$) exists if $w_{ij} = c_{ij} - c_{ji}$ is negative and its corresponding unweighted edge is $a_{ji} = 1$; in that case we set and $w_{ij} = 0$ and $a_{ij} = 0$.

Step 4 Construction of the Network: A global net food calorie trade network is constructed, where each node represents a country and the edge direction indicates the flow of net food calorie trade between country pairs. In the context of a weighted network, the weight of each edge quantifies the net food calorie trade flow, while in an unweighted network, each edge indicates the presence of net caloric trade flow without specific quantification.

This network provides a comprehensive representation of the global food trade dynamics, taking into account both the direction and magnitude of the caloric flows between countries.

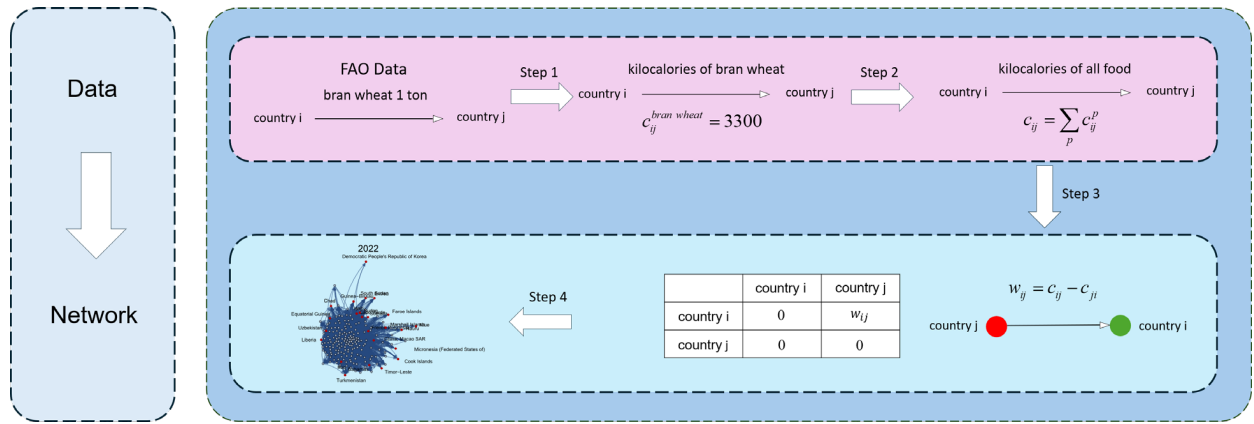

**Fig. 1 | Schematic representation of the methodological framework employed in this study for constructing the global net food calorie trade networks. The process is outlined in Section 2.3: (1) Conversion of trade data to caloric content, (2) Aggregation of export records, (3) Calculation of net caloric export, and (4) Construction of the network.**

2.4 Network measures

In this study, we employ a range of network measures, including both weighted and unweighted metrics, to perform a thorough analysis of global net food calorie trade networks (Boccaletti et al., 2006; Costa et al., 2007; Kwapień & Drożdż, 2012; M. E. J. Newman, 2010). This methodological approach enables us to capture not only the structural configuration of the network but also the magnitude of trade relationships in terms of caloric content. By integrating these measures, we can discern the intensity and directionality of caloric exchanges between countries, thereby providing a more detailed and nuanced understanding of global food trade dynamics.

2.4.1 Degree

In the complex network analysis, the unweighted degree of a given node refers to the number of edges connected to it (M. E. J. Newman, 2010). For a node i, the unweighted in-degree $k_i^{in}$ is determined by the sum of ingoing edges to i and is calculated as:

$$k_i^{in} = \sum_j a_{ij}$$

where $a_{ij}$ represents availability of net caloric flow from country j to country i. Similarly, the unweighted out-degree of a node is calculated as:

$$k_i^{out} = \sum_j a_{ji}$$

In weighted networks, the concept of "weighted degree" or "node strength" representing the aggregate of the weights of all edges that are connected to a given node (M. E. J. Newman, 2010). This concept is particularly relevant in weighted networks, where edges have different levels of significance or intensity rather than being simply present or absent. In directed networks, weighted in-degree of a node is computed as the cumulative sum of the weights of all incoming edges to that node, mathematically represented as:

$$s_i^{in} = \sum_j w_{ij}$$

where $w_{ij}$ represents the net caloric flow from country j to country i. The weighted in-degree denotes the total calorie imports from all other countries, so a country with a high weighted out-degree might be identified as a significant exporter. Similarly, the weighted out-degree of a node is calculated as the aggregate sum of the weights of all outgoing edges from that node, mathematically represented as:

$$s_i^{out} = \sum_j w_{ji}$$

The weighted out-degree indicates the total calorie exports to all other countries, so a country with a high weighted in-degree might be characterized as a major importer.

2.4.2 Network connectivity

Network connectivity is a fundamental aspect of complex network analysis, crucial for understanding the robustness, efficiency, and overall functionality of various systems. It refers to the degree to which nodes (or vertices) within the network are connected to each other. In highly connected systems, a shock or disturbance can propagate more easily through the network, potentially affecting all its components. The network connectivity is defined as follows (M. E. J. Newman, 2010):

$$c = \frac{L}{N(N-1)}$$

where $N$ and $L$ are the number of node and edge, respectively. This formula provides a measure of the density of connections within the network, offering insights into its structural properties and potential for systemic risk.

2.4.3 Network heterogeneity

Heterogeneity in complex networks pertains to the variation and diversity in the structural attributes of the nodes and edges within the network. A key metric for assessing heterogeneity is the distribution of in-degree and out-degree across nodes. In homogeneous networks, a strong positive correlation typically exists between in-degree and out-degree, whereas in heterogeneous networks, this correlation tends to be negative. In this study, we employ the indicator designed to quantify the degree of heterogeneity in our previous work (Chengyi Tu et al., 2019):

$$h = \frac{\langle | k^{in} - k^{out} | \rangle}{\langle k \rangle} \in [0,1]$$

where the numerator is the average of the absolute value of the difference between unweighted in-degree $k^{in}$ and out-degree $k^{out}$ of each node. If the network is homogeneous, the absolute value of each node is positive small; if heterogenous, it is positive large. To normalize the value so that it ranges between 0 and 1, we divide it by the mean network degree $\langle k \rangle$.

For weighted networks, we extend this measure to capture weighted heterogeneity:

$$h^w = \frac{\langle | s^{in} - s^{out} | \rangle}{\langle s \rangle} \in [0,1]$$

where the numerator is the average of the absolute value of the difference between weighted in-degree $s^{in}$ and out-degree $s^{out}$ of each node and $\langle s \rangle$ is the mean network degree.

If the network heterogeneity approximates to 0, the network is completely homogeneous, meaning all nodes have similar connectivity. If the network heterogeneity approximates to 1, the network is completely heterogenous; for example, in a bipartite network where one half of the nodes have a fixed in-degree and zero out-degree, and the other half have a fixed out-degree and zero in-degree. This definition normalizes the two extreme cases as boundaries and does not account for other structural properties of the network.

2.4.4 Network modularity

Modularity quantifies the extent to which a network can be partitioned into distinct modules, also referred to as communities or clusters. This property is characteristic of networks that contain sub-networks of nodes exhibiting stronger internal connectivity compared to the average connectivity across the entire network (Grilli et al., 2016; May et al., 2008; Scheffer et al., 2012; Stouffer & Bascompte, 2011; Chengyi Tu et al., 2019). Consider an unweighted network where each sub-network $g_i$ comprises nodes with internal edges $m_l$ and internal connectivity $c_l$. The number of edges connecting the nodes from different sub-networks is denoted as $m_g$ and its connectivity, or global connectivity, is $c_g$. The total number of edges in the network $g$ is $m$ and the total connectivity is $c$. Let us denote by $g_i$ the sub-network of node $i$, which is an integer $\{1 \ldots n_g\}$, with $n_g$ being the total number of sub-networks, the modularity $Q$ is defined as follows (M. E. Newman, 2003):

$$Q = \frac{1}{2E} \sum_{ij} \left( a_{ij} - \frac{k_i k_j}{2E} \right) \Delta(g_i, g_j)$$

where $a_{ij}$ represents availability of net caloric flow from country j to country i, $E$ is the total number of edges in the network, $k_i$ and $k_j$ are the unweighted degrees of nodes i and j, respectively, $\Delta(x,y)$ is the Kronecker delta.

For weighted networks, the modularity is defined as (M. E. Newman & Girvan, 2004):

$$Q^w = \frac{1}{2W} \sum_{ij} \left( w_{ij} - \frac{s_i s_j}{2W} \right) \Delta(g_i, g_j)$$

where $w_{ij}$ represents strength of net caloric flow from country j to country i, $W$ is total weight of all edges in the network, $s_i$ and $s_j$ are the weighted degrees of nodes i and j, respectively. This formulation accounts for both the existence and the intensity of connections within and across communities.

A modularity indicates that there are more edges within the same sub-network than would be expected by chance, while a negative modularity suggests the opposite. Networks with high modularity exhibit dense connections within modules and sparse connections between different modules.

2.4.5 Node correlation similarity

Node correlation similarity emerges as a pivotal metric, quantifying the likeness between nodes based on their connections and attributes (Barabasi & Albert, 1999; M. E. J. Newman, 2010; Watts & Strogatz, 1998). This measure is instrumental in deciphering the structural properties and dynamics of complex networks, thereby shedding light on the interrelationships between nodes within the network. Node similarity, in essence, quantifies the degree to which two nodes share common characteristics or connections. This can be assessed through various factors, including the number of shared neighbors, node attributes, or the overall network structure. In unweighted networks, we utilize the Pearson Correlation Coefficient as a measure for edge. This coefficient gauges the linear correlation between the degrees of connected nodes, mathematically expressed as (M. E. J. Newman, 2010):

$$\bar{r} = \frac{\sum_{ij} r_{ij}}{N(N-1)} = \frac{\sum_j (a_{ij} - k^{in})(a_{ji} - k^{out})}{\sqrt{\sum_j (a_{ij} - k^{in})^2 \sum_j (a_{ji} - k^{out})^2}}{N(N-1)}$$

For weighted networks, the node correlation similarity is defined as (M. E. J. Newman, 2010):

$$\overline{r^w} = \frac{\dfrac{\sum_k (w_{ik} - s^{in})(w_{jk} - w^{out})}{\sqrt{\sum_k (w_{ik} - s^{in})^2 \sum_k (w_{jk} - s^{out})^2}}}{N(N-1)}$$

In the specific context of the global net food calorie trade networks, node correlation similarity refers to the statistical and topological similarity of nodes (countries). This measure provides a quantitative assessment of the extent to which two countries exhibit similar patterns in their food calorie trade networks. It allows us to identify countries with similar roles and positions within the network, which can, in turn, inform our understanding of global food security dynamics.

**3 Results**

3.1 Evolution of weighted degree

This study conducts a comprehensive analysis of the global net food calorie trade flows—specifically the weighted degree of trade networks—among countries from 1986 to 2022 (see Fig. 2). This analysis categorizes the timeframe into four distinct phases: (1) the pre-Soviet Union dissolution period (1986-1991), (2) the post-Soviet era until China's accession to the World Trade Organization (1992-2001), (3) the period from 2002 to 2012, characterized by the global impacts before the Arab Spring, and (4) the interval from 2012 to the present. The weighted degrees within the global net food calorie trade networks reveals considerable heterogeneity across countries. Certain countries display a high weighted in-degree, indicative of substantial import net food calorie, coupled with a low weighted out-degree, reflecting minimal export net food calorie. Conversely, other countries exhibit the opposite pattern. Notably, no country simultaneously demonstrates both a high weighted in-degree and out-degree, suggesting a tendency for countries to specialize predominantly in either importing or exporting activities.

The majority of countries are characterized by both low weighted in-degree and out-degree, signifying a relatively balanced net food calorie (see Fig. 2 a). A subset of countries is distinguished by a minimal weighted in-degree and a zero weighted out-degree, indicating a primary role as importers with negligible export net food calorie. Further, it suggests that the country has limited trade dependence or relies on imports from only a few countries. As a result, the country's resilience to fluctuations in food trade is relatively low. The network also encompasses numerous loosely connected peripheral countries that engage in minimal export activities and do not import. In contrast, a select group of countries is responsible for exporting net food calorie, with some exhibiting pronounced export activity. These observations highlight the complex nature of global food trade. The patterns observed remain consistent over time, with both the weighted in-degree and out-degree generally increasing, indicative of an expansion in global net food calorie trade throughout the observed period. Moreover, the correlation between the weighted out-degree and in-degree remains relatively stable throughout the observation period (see Fig. 2 b).

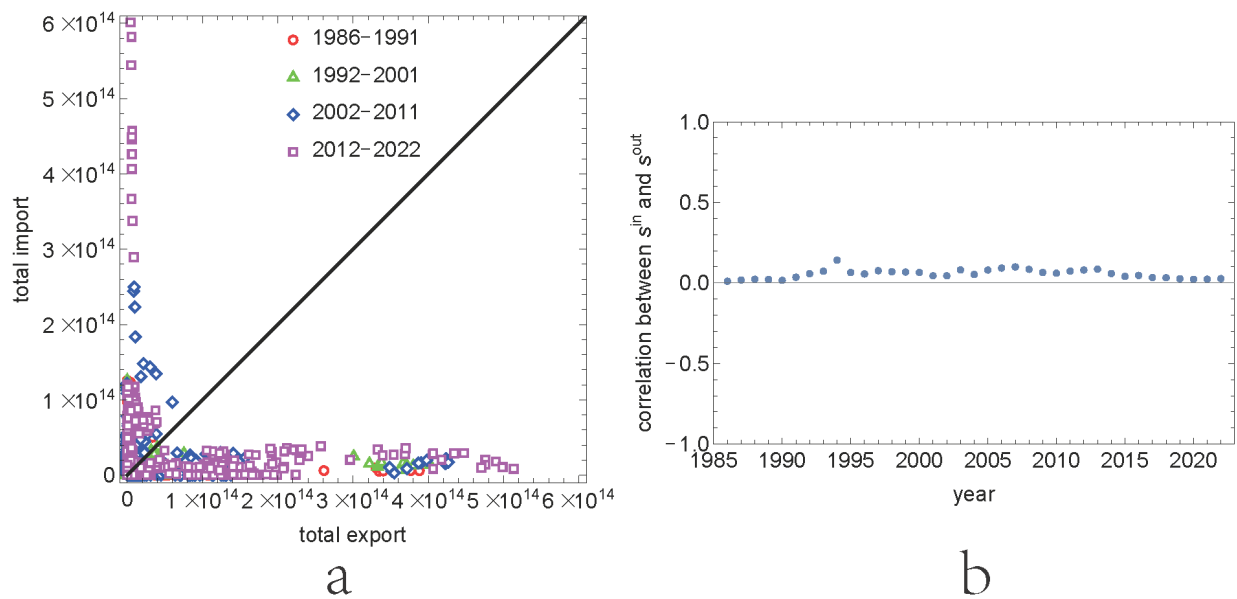

**Fig. 2 | Relationship between net weighted out-degree and in-degree from 1986 to 2022. (a) Scatter plot depicting the distribution of weighted out-degree versus in-degree for each country across the years 1986 to 2022. Each data point represents a specific country in a given year, highlighting the variation in the relationship between out-degree and in-degree over time. (b) Correlation between weighted out-degree and in-degree throughout the observed period. The results indicate that the relationship between net importation and exportation remains largely uncorrelated and exhibits minimal variation over the years.**

3.2 Highest export and import activities

In an effort to comprehensively identify the countries with the highest exports (weighted out-degree) and imports (weighted in-degree), Fig. 3 delineates the top five countries with the greatest weighted out-degree and in-degree for each year. Major exporters (i.e., countries with the highest weighted out-degrees) include Brazil, the United States of America, Argentina, Ukraine, Indonesia, Australia, Canada, France, the Russian Federation, and Malaysia, which, altogether, account for about 60% of global food calorie trade (see Fig. 3 a). Conversely, the major importers (i.e., countries with the highest weighted in-degree) encompass mainland China, Mexico, Japan, the Republic of Korea, Spain, Egypt, Italy, Bangladesh, the Netherlands, and Saudi Arabia (see Fig. 3 b).

A notable observation is that countries with limited land-based agricultural resources (relative to their populations), such as China, India, and Japan, consistently emerge as major importers. In contrast, countries endowed with abundant fertile land resources, such as the USA, Russia, and Canada, remain dominant exporters. The export volume, quantified in kilocalories, exhibits variability among different countries. Notably, Brazil has experienced a notable expansion in its net exports, with a fourfold increase, reaching approximately $4.1 \times 10^{14}$, while United States has maintained a high and stable level of net exports. However, within the cohort of importers, after 2005 China becomes particularly prominent, importing a volume that exceeds the aggregate import volumes of the other major importing countries. Notably, China's net imports and the rate of increase in its import volume post-2005 have been strikingly significant. The observed shifts in net imports and exports can primarily be attributed to significant increases in the volumes of both exports and imports, respectively. Throughout the study period, the composition of the leading import and export countries has remained relatively stable, indicating a persistent equilibrium in global food trade dynamics. This stability underscores the enduring influence of geographical and economic factors on global food trade patterns, highlighting the interplay between resource availability and trade activities.

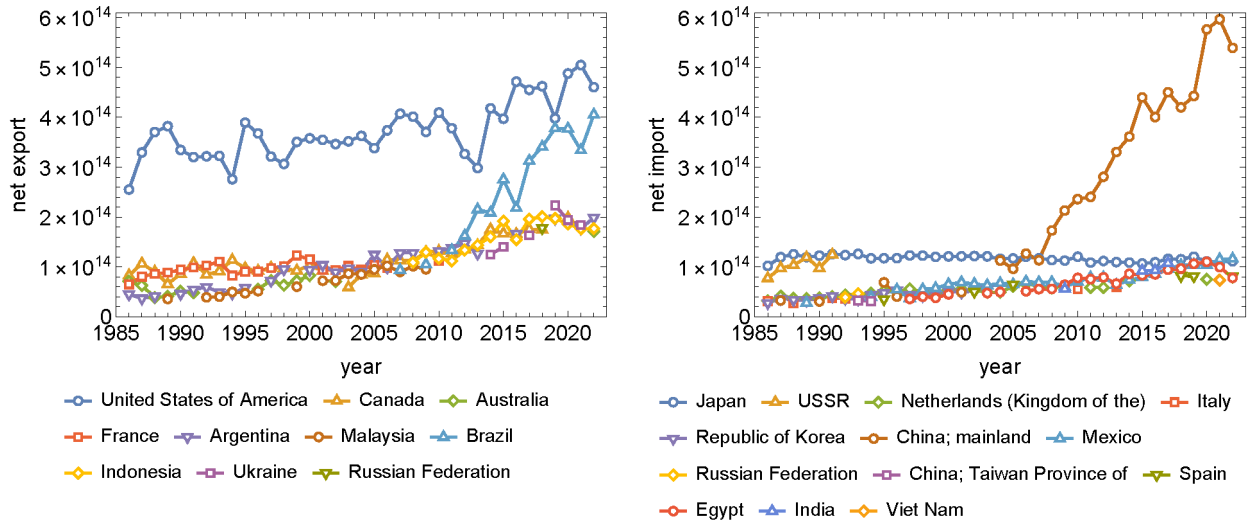

**Fig. 3 | The five largest export and import countries. (a) The five largest global net food calorie exporters for each year from 1986 to 2022. (b) The five largest global net food calorie importers for each year from 1986 to 2022.**

3.3 Principal actors for peripheral countries

In addition to examining the principal actors within the global net food calorie trade networks, our study also scrutinizes the so-called "peripheral" countries. These countries are characterized by a weighted out-degree of zero kilocalories, signifying their non-participation in net food calorie exports. The proportion of zero-export countries within the global net food calorie trade network (see Fig. 4a) shows a general trend of decreasing zero-export fractions over the observed periods. This trend can be delineated into three distinct stages: From 1986 to 1996, the proportion of zero-export countries was relatively high, with an average value of approximately 55%, as indicated by net food calorie trade flows. This period is characterized by a significant number of countries not participating in the export of food calories. From 2001 to 2013, there was a notable reduction in the fraction of zero-export countries, with the average value decreasing to around 30%. This suggests an increasing number of countries began engaging in the export of food calories during this period. From 2014 to 2022, the fraction of zero-export countries further declined to an average value of about 10%. This stage reflects a substantial integration of countries into the global food calorie trade network, with a minimal number of countries remaining as zero-exporters. Overall, the data presented in Fig. 4a highlight a progressive trend towards greater participation in the global net food calorie trade, as evidenced by the decreasing fraction of zero-export countries across the three identified stages.

Fig. 4b provides a visualization of the 2022 global food trade network, with these "peripheral" countries highlighted for emphasis. The countries identified in this category include the Bahamas, Cabo Verde, Chad, Equatorial Guinea, Haiti, Iraq, the Democratic People's Republic of Korea, Liberia, the Marshall Islands, China; Macao SAR, the Maldives, Micronesia (Federated States of), Nauru, Niue, Guinea-Bissau, Timor-Leste, Eritrea, Somalia, Turkmenistan, Tokelau, Uzbekistan, Sudan, and South Sudan. These countries are completely dependent on international food trade.

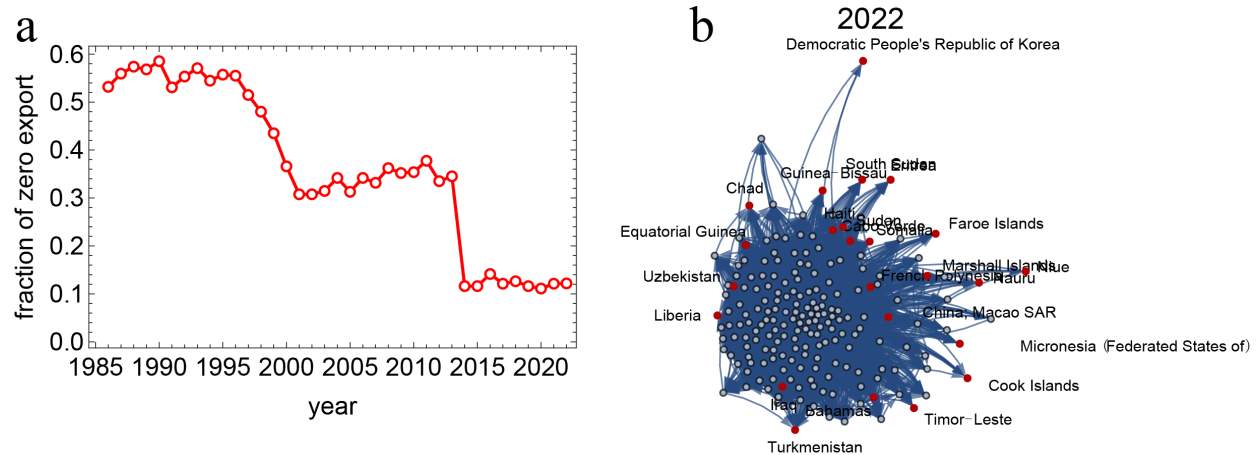

**Fig. 4 | Peripheral countries within the global net food calorie trade networks. (a) Proportion of peripheral countries from 1986 to 2022. (b) Network visualization for the year 2022, highlighting peripheral countries.**

3.4 Evolution of network structure

Finally, we employ a series of measures to scrutinize the structural dynamics of the global net food calorie trade network across both weighted and unweighted network. The temporal evolution of key network indicators, including connectivity, heterogeneity, modularity, and node correlation similarity, spanning the period from 1986 to 2022 (see Fig. 5) provides an important description of the way the network structure has been changing in the last few decades.

Throughout the entire observation period, the connectivity within the global net food calorie trade network exhibits an increase from 0.1 to 0.2. It is important to note that, given the structure of this network, each pair of nodes is connected by a single edge. Consequently, this increase in connectivity is significant and should not be considered low because the net flow between two countries accounts for exchanges of multiple commodities in both directions. This observation aligns with the findings of our previous research (Chengyi Tu et al., 2019), further underscoring the ongoing globalization of food and evolving nature of global food trade networks.

During the observation period, the global net food calorie trade networks exhibit considerable heterogeneity, both in weighted and unweighted forms. The weighted networks consistently maintained a value close to 0.8, while the unweighted networks approximated 0.6. This high and stable level of heterogeneity indicates that the distribution of trade connections among countries remained markedly uneven, with a negative correlation observed between the number and strength of inflow and outflow links across countries.

A significant change in the modularity of the global food calorie trade networks is observed, particularly in the weighted version of the network. Modularity increased from approximately 0.28 to around 0.36, followed by a period of stabilization at this higher value. This increase indicates a growing tendency towards the formation of distinct clusters or communities within the network, suggesting that trade interactions became more compartmentalized (i.e., 'regionalized') over time. The stabilization at a modularity value of 0.36 might reflect a new equilibrium state, where the network's structure is characterized by a higher degree of community organization. In contrast, the modularity of the unweighted version of the global food calorie trade networks showed a marked decline, decreasing from about 0.14 to 0.08 over the same period. This contrasting trend in the unweighted network suggests a reduced level of structural organization, with less clear delineation of communities or clusters in the absence of trade weight considerations.

The weighted global food calorie trade network displayed a noticeable decline in node correlation similarity, decreasing from approximately 0.25 to around 0.15. This downward trend indicates a reduction in the similarity of trade patterns among nodes, suggesting that the network became less homogeneous over time in agreement with the observed changes in trade heterogeneity. The decreasing node correlation similarity reflects a diversification in the trade relationships, where countries' trading partners became more distinct from one another. In contrast, the unweighted version of the network exhibited a relatively stable node correlation similarity, maintaining a value around 0.3 throughout the period, indicating a less pronounced change in the relative distribution of trade partnerships. This discrepancy between the weighted and unweighted networks further highlights the importance of considering trade intensity in understanding the dynamics of global trade networks.

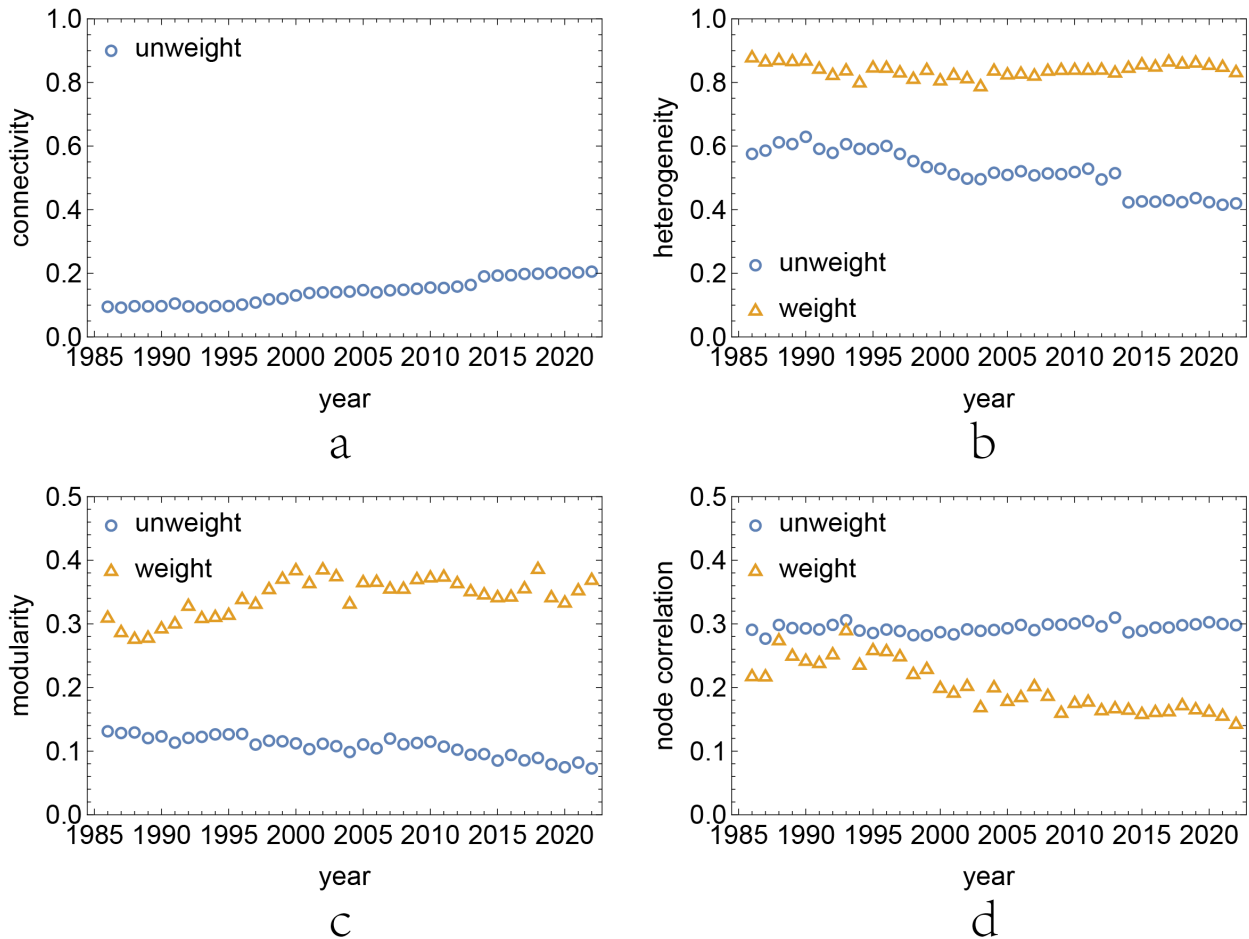

**Fig. 5 | Unweighted and weighted measures of the global net food calorie trade networks. (a) network connectivity, (b) network heterogeneity, (c) network modularity, and (d) node correlation similarity, analyzed over the period from 1986 to 2022.**

**4 Discussion**

This study analyzes global food trade by converting the weight of traded food items into a unified measure of caloric content. This methodology allows for a more nuanced understanding of global food trade dynamics, emphasizing the nutritional value of traded goods rather than merely their volume. By constructing a directed, weighted network of net caloric flows between countries, we provide a comprehensive representation of global food trade that highlights the importance of nutritional considerations in discussions of food security and trade policies. To our knowledge, this is the first study to explore net caloric flow specifically. Previous works predominantly centered on actual caloric flow. Consequently, their analyses of network metrics may yield different results due to this distinction in focus.

The results reveal significant heterogeneity in the trade dynamics among countries (Duan et al., 2021; Dupas et al., 2019). Certain countries, such as Brazil and the United States, emerge as major exporters of net food calories, while others, including China and Japan, are prominent importers. The findings highlight a considerable diversity in the trade dynamics among countries. Countries with abundant land resources, such as Brazil and the United States, are typically net exporters, in agreement with previous studies (Carr et al., 2012; D'Odorico et al., 2014; Porkka et al., 2013). These countries have vast agricultural lands that allow for large-scale food production, enabling them to export a surplus of food calories to other countries. Their geographical location, favorable climate, and advanced agricultural technologies further contribute to their high productivity. On the other hand, countries with more limited cultivable land resources (relative to their population), such as China and Japan, are typically net importers. Despite having strong economies, these countries do not have enough land (and water) to meet their food demand with the domestic production. As a result, they rely on food imports to meet their population's dietary needs. This pattern of trade reflects the global division of labor in food production and the interdependence of countries in ensuring food security. It also highlights the importance of international trade in balancing the global food supply and demand. However, this pattern also has its vulnerabilities. For instance, any disruption in the food supply from exporting countries can have significant impacts on the food security of importing countries. Similarly, changes in trade policies or global food prices can also affect the affordability and accessibility of food in different countries.

Our analysis also highlights the presence of peripheral countries that do not contribute to net food calorie exports (Friedmann, 2021; Grodzicki & Geodecki, 2016). Their agricultural sectors may be insufficiently developed because of their limited availability of productive agricultural land, water, economic or institutional capacity, making it difficult for them to produce enough food to meet their population's needs. As a result, they rely heavily on international food trade to supplement their domestic food production. However, their dependence on food imports presents its own set of challenges. Fluctuations in global food prices, changes in trade policies, or disruptions in supply chains can have significant impacts on these countries' food security (Fader et al., 2011). Changes in the amount of food these peripheral countries are able to import can have a disproportionate impact on their food security status. Moreover, these countries often lack the bargaining power to negotiate favorable trade terms, which can further exacerbate their food insecurity (Margulis, 2018; Zurek et al., 2022). They may also face challenges in accessing nutritious food, as the food that is available for import may not always align with the nutritional needs of their population.

Furthermore, the study of temporal trends in network metrics (M. E. J. Newman, 2010), including network connectivity, heterogeneity, modularity, and node correlation similarity, provides valuable insights into the structural dynamics of global net food calorie trade networks. In the specific context of global net food calorie trade networks, connectivity refers to the degree in which different countries or regions are linked through the exchange of food calories. This concept encompasses the structural properties of the network, such as the number and strength of trade links, the directionality of calorie flows, and the overall network topology.

The observed increase in connectivity indicates a greater level of integration within the global net food calorie trade network, resulting in improved pathways for the redistribution of food calories. This enhancement is particularly significant in the context of global food security, as higher connectivity can alleviate the effects of localized food shortages by enabling more efficient and effective redistribution of resources. The consistent heterogeneity underscores the intricate and multifaceted nature of global food trade dynamics, where certain countries assume more central roles while others remain peripheral. This network property plays as crucial role in determining the stability and resilience of the global food system. In fact, in heterogeneous networks an increase in connectivity is associated with a decrease in resilience (Chengyi Tu et al., 2019).

In the specific context of the global net food calorie trade networks, modularity plays a crucial role in understanding the structural organization and resilience of these networks. It refers to the degree to which the network can be divided into distinct modules or communities, where each module represents a group of countries with stronger internal trade connections compared to their external trade connections with other modules. High modularity in global food trade networks indicates that countries within the same module trade more intensively with each other than with countries outside their module. This can reveal important insights into regional trade patterns, dependencies, and potential vulnerabilities. For instance, a high modularity score might suggest that certain regions are more self-sufficient or have stronger intra-regional trade ties, which could enhance their resilience to global market fluctuations or disruptions. The observed increase in modularity is particularly significant as it suggests alterations in the underlying trade patterns, potentially driven by geopolitical, economic, or policy factors that influence the segmentation of global food trade. This shift in modularity indicates a growing tendency towards the formation of distinct clusters or communities within the network, reflecting changes in trade interactions over time. These findings contrast with prior studies that reported a temporal decline in network modularity within the global food network. For instance, Tu et al. (Chengyi Tu et al., 2019) derive network weights directly from food trade volumes (in tons), employing an unweighted modularity measure. Conversely, D'Odorico et al. (D'Odorico et al., 2014; D'Odorico et al., 2012) analyze actual calorie flows, utilizing a weighted modularity measure. In this study, we adopt net calorie flows as the basis for network weights and similarly apply a weighted modularity measure. Interestingly, in heterogeneous networks an increase in modularity is expected to enhance the resilience of the food system (Chengyi Tu et al., 2019).

Additionally, the decline in node correlation similarity from approximately 0.25 to around 0.15 indicates a reduction in the similarity of trade patterns among nodes, suggesting that the network became less homogeneous. This diversification in trade relationships could be attributed to various factors, including changes in trade policies, economic conditions, or geopolitical influences that drive countries to diversify their trade connections. These findings are essential for comprehending the resilience and vulnerability of the global food trade network, as they highlight the dynamic nature of global food trade and its implications for global food security.

Recent work has investigated the way the network structure may affect the resilience of global food system (Chengyi Tu et al., 2019). If we envision the resources available for food production at each node as being prone to critical transitions as the demand (i.e., harvest rate) exceeds a critical threshold, the global system also exhibits a critical transition at a threshold that depends on the network structure, thus affecting the resilience of the global food system. In heterogeneous networks the resilience decreases with an increase in network connectivity and a decrease in modularity. Globalization typically entails both a growing connectivity and a decrease in regionalization (i.e., modularity), thus potentially favoring a critical transition to a resource depleted state. As noted earlier, our analyses show an increase in modularity, which in a heterogenous network should correspond to an increase in resilience, thus partly counteracting the effect of the ongoing globalization (increasing connectivity) trends.

**Acknowledgments**

This work was supported by the Natural Science Foundation of Zhejiang Sci-Tech University (grant number 22092034-Y), Social Science Foundation of Zhejiang Province (grant number 24096010-G) and National Natural Science Foundation of China (grant number 42001021).

**Data Availability Statement**

The ready-to-use notebook codes and data to reproduce the results presented in the current study are available in OSF with the access code rgft5 (https://osf.io/rgft5/).

**References**

Alexandratos, N., & Bruinsma, J. (2012). World agriculture towards 2030/2050: the 2012 revision.
Ardan, M. C., Marzaman, A. P., & Deni, S. (2023). The Russian-Ukrainian War Effects on Global Food Trade: A Review. *J. Political Sci. Int. Relat, 6*, 54.

Barabasi, A.-L., & Albert, R. (1999). Emergence of Scaling in Random Networks. *Science, 286*(15), 509-512.

Barigozzi, M., Fagiolo, G., & Mangioni, G. (2011). Identifying the community structure of the international-trade multi-network. *Physica A: Statistical Mechanics and its Applications, 390*(11), 2051-2066.

Boakes, E. H., Dalin, C., Etard, A., & Newbold, T. (2024). Impacts of the global food system on terrestrial biodiversity from land use and climate change. *Nat Commun, 15*(1), 5750.

Boccaletti, S., Latora, V., Moreno, Y., Chavez, M., & Hwang, D. (2006). Complex networks: Structure and dynamics. *Physics Reports, 424*(4-5), 175-308.

Carlsen, L., & Bruggemann, R. (2022). The 17 United Nations' sustainable development goals: A status by 2020. *International Journal of Sustainable Development & World Ecology, 29*(3), 219-229.

Carr, J. A., D'Odorico, P., Laio, F., & Ridolfi, L. (2012). On the temporal variability of the virtual water network. *Geophysical Research Letters, 39*(6).

Chaudhuri, A., Srivastava, S. K., Srivastava, R. K., & Parveen, Z. (2016). Risk propagation and its impact on performance in food processing supply chain: A fuzzy interpretive structural modeling based approach. *Journal of Modelling in Management, 11*(2), 660-693.

Chung, M. G., & Liu, J. G. (2022). International food trade benefits biodiversity and food security in low-income countries. *Nature Food, 3*(5), 349-+. <Go to ISI>://WOS:000799082000002

Costa, L. d. F., Rodrigues, F. A., Travieso, G., & Villas Boas, P. R. (2007). Characterization of complex networks: A survey of measurements. *Advances in Physics, 56*(1), 167-242.

D'Odorico, P., Carr, J. A., Laio, F., Ridolfi, L., & Vandoni, S. (2014). Feeding humanity through global food trade. *Earth's Future, 2*(9), 458-469.

D'Odorico, P., Davis, K. F., Rosa, L., Carr, J. A., Chiarelli, D., Dell'Angelo, J., et al. (2018). The global food‐energy‐water nexus. *Reviews of Geophysics, 56*(3), 456-531.

D'Odorico, P., Carr, J., Dalin, C., Dell'Angelo, J., Konar, M., Laio, F., et al. (2019). Global virtual water trade and the hydrological cycle: patterns, drivers, and socio-environmental impacts. *Environmental Research Letters, 14*(5), 053001.

D'Odorico, P., Carr, J., Laio, F., & Ridolfi, L. (2012). Spatial organization and drivers of the virtual water trade: A community-structure analysis. *Environmental Research Letters, 7*(3), 034007.

Dalin, C., Konar, M., Hanasaki, N., Rinaldo, A., & Rodriguez-Iturbe, I. (2012). Evolution of the global virtual water trade network. *Proceedings of the National Academy of Sciences, 109*(16), 5989-5994.

Dalin, C., & Outhwaite, C. L. (2019). Impacts of global food systems on biodiversity and water: the vision of two reports and future aims. *One Earth, 1*(3), 298-302.

Davis, K. F., Downs, S., & Gephart, J. A. (2021). Towards food supply chain resilience to environmental shocks. *Nature Food, 2*(1). <Go to ISI>://WOS:000603467600001

Duan, J., Nie, C., Wang, Y., Yan, D., & Xiong, W. (2021). Research on global grain trade network pattern and its driving factors. *Sustainability, 14*(1), 245.

Dupas, M.-C., Halloy, J., & Chatzimpiros, P. (2019). Time dynamics and invariant subnetwork structures in the world cereals trade network. *PLoS ONE, 14*(5), e0216318.

Ercsey-Ravasz, M., Toroczkai, Z., Lakner, Z., & Baranyi, J. (2012). Complexity of the international agro-food trade network and its impact on food safety. *PLoS ONE, 7*(5), e37810.

Fader, M., Gerten, D., Thammer, M., Heinke, J., Lotze-Campen, H., Lucht, W., & Cramer, W. (2011). Internal and external green-blue agricultural water footprints of nations, and related water and land savings through trade. *Hydrology and Earth System Sciences, 15*(5), 1641-1660.

Falcon, W. P., Naylor, R. L., & Shankar, N. D. (2022). Rethinking global food demand for 2050. *Population and Development Review, 48*(4), 921-957.

Friedmann, H. (2021). Changes in the international division of labor: agri-food complexes and export agriculture. In *Towards a new political economy of agriculture* (pp. 65-93): Routledge.

Giller, K. E., Delaune, T., Silva, J. V., Descheemaeker, K., van de Ven, G., Schut, A. G. T., et al. (2021). The future of farming: Who will produce our food? *Food Security, 13*(5), 1073-1099.

Gomez, M., Mejia, A., Ruddell, B. L., & Rushforth, R. R. (2021). Supply chain diversity buffers cities against food shocks. *Nature, 595*(7866), 250-254. Article. <Go to ISI>://WOS:000671377900013

Grilli, J., Rogers, T., & Allesina, S. (2016). Modularity and stability in ecological communities. *Nat Commun, 7*, 12031. https://www.ncbi.nlm.nih.gov/pubmed/27337386

Grodzicki, M. J., & Geodecki, T. (2016). New dimensions of core-periphery relations in an economically integrated Europe: The role of global value chains. *Eastern European Economics, 54*(5), 377-404.

Halpern, B. S., Frazier, M., Verstaen, J., Rayner, P. E., Clawson, G., Blanchard, J. L., et al. (2022). The environmental footprint of global food production. *Nature Sustainability*. <Go to ISI>://WOS:000871310200001

Hicks, C. C., Gephart, J. A., Koehn, J. Z., Nakayama, S., Payne, H. J., Allison, E. H., et al. (2022). Rights and representation support justice across aquatic food systems. *Nature Food, 3*(10), 851-861. <Go to ISI>://WOS:000869596700003

Jia, S. S., Gibson, A. A., Ding, D., Allman-Farinelli, M., Phongsavan, P., Redfern, J., & Partridge, S. R. (2022). Perspective: are online food delivery services emerging as another obstacle to achieving the 2030 United Nations sustainable development goals? *Frontiers in Nutrition, 9*, 858475.

Karakoc, D. B., & Konar, M. (2021). A complex network framework for the efficiency and resilience trade-off in global food trade. *Environmental Research Letters, 16*(10). <Go to ISI>://WOS:000700767500001

Karakoc, D. B., Konar, M., Puma, M. J., & Varshney, L. R. (2023). Structural chokepoints determine the resilience of agri-food supply chains in the United States. *Nature Food, 4*(7), 607-615.

Karakoc, D. B., Wang, J., & Konar, M. (2022). Food flows between counties in the United States from 2007 to 2017. *Environmental Research Letters, 17*(3), 034035.

Konar, M., Dalin, C., Suweis, S., Hanasaki, N., Rinaldo, A., & Rodriguez‐Iturbe, I. (2011). Water for food: The global virtual water trade network. *Water Resources Research, 47*(5).

Konar, M., Lin, X., Ruddell, B., & Sivapalan, M. (2018). Scaling properties of food flow networks. *PLoS ONE, 13*(7), e0199498.

Kreibich, H., Van Loon, A. F., Schroter, K., Ward, P. J., Mazzoleni, M., Sairam, N., et al. (2022). The challenge of unprecedented floods and droughts in risk management. *Nature, 608*(7921), 80-+. <Go to ISI>://WOS:000835655400012

Kwapień, J., & Drożdż, S. (2012). Physical approach to complex systems. *Physics Reports, 515*(3-4), 115-226.

Laborde, D., Martin, W., Swinnen, J., & Vos, R. (2020). COVID-19 risks to global food security. *Science, 369*(6503), 500-502.

Marchand, P., Carr, J. A., Dell'Angelo, J., Fader, M., Gephart, J. A., Kummu, M., et al. (2016). Reserves and trade jointly determine exposure to food supply shocks. *Environmental Research Letters, 11*(9), 095009.

Margulis, M. E. (2018). Negotiating from the margins: how the UN shapes the rules of the WTO. *REVIEW OF INTERNATIONAL POLITICAL ECONOMY, 25*(3), 364-391.

May, R. M., Levin, S. A., & Sugihara, G. (2008). Complex systems: Ecology for bankers. *Nature, 451*(7181), 893.

Mehrabi, Z., Ignaciuk, A., Levers, C., Delzeit, R., Braich, G., Bajaj, K., et al. (2022). Research priorities for global food security under extreme events. *One Earth, 5*(7), 756-+. <Go to ISI>://WOS:000838639500007

Mekonnen, M. M., Kebede, M. M., Demeke, B. W., Carr, J. A., Chapagain, A., Dalin, C., et al. (2024). Trends and environmental impacts of virtual water trade. *Nature Reviews Earth & Environment*, 1-16.

Muller, M. F., Penny, G., Niles, M. T., Ricciardi, V., Chiarelli, D. D., Davis, K. F., et al. (2021). Impact of transnational land acquisitions on local food security and dietary diversity. *Proc Natl Acad Sci U S A, 118*(4). <Go to ISI>://WOS:000612945500065

Nakai, J. (2018). Food and Agriculture Organization of the United Nations and the sustainable development goals. *Sustainable development, 22*, 1-450.

Newman, M. E. (2003). Mixing patterns in networks. *Physical Review E, 67*(2), 026126.

Newman, M. E., & Girvan, M. (2004). Finding and evaluating community structure in networks. *Physical Review E, 69*(2), 026113.

Newman, M. E. J. (2010). *Networks: An Introduction*. United States: Oxford University Press Inc., New York

Oluwole, O., Ibidapo, O., Arowosola, T., Raji, F., Zandonadi, R. P., Alasqah, I., et al. (2023). Sustainable transformation agenda for enhanced global food and nutrition security: a narrative review. *Frontiers in Nutrition, 10*, 1226538.

Organization, S. D. o. t. F. a. A. (2024a). Nutritive Factors. https://www.fao.org/economic/the-statistics-division-ess/publications-studies/publications/nutritive-factors/en/

Organization, S. D. o. t. F. a. A. (2024b). Statistics Division of the Food and Agriculture Organization. Retrieved from https://www.fao.org/faostat/en/#home

Porkka, M., Kummu, M., Siebert, S., & Varis, O. (2013). From food insufficiency towards trade dependency: a historical analysis of global food availability. *PLoS ONE, 8*(12), e82714.

Read, Q. D., Brown, S., Cuéllar, A. D., Finn, S. M., Gephart, J. A., Marston, L. T., et al. (2020). Assessing the environmental impacts of halving food loss and waste along the food supply chain. *Science of the Total Environment, 712*, 136255.

Scheffer, M., Carpenter, S. R., Lenton, T. M., Bascompte, J., Brock, W., Dakos, V., et al. (2012). Anticipating critical transitions. *Science, 338*(6105), 344-348.

Silvestrini, M. M., Smith, N. W., & Sarti, F. M. (2023). Evolution of global food trade network and its effects on population nutritional status. *Current Research in Food Science, 6*, 100517.

Springmann, M., Kennard, H., & Dalin, C. (2023). International food trade contributes to dietary risks and mortality at global, regional and national levels. *Nature Food, 4*(10), 886-893.

Stein, A. J., & Santini, F. (2022). The sustainability of "local" food: A review for policy-makers. *Review of Agricultural, Food and Environmental Studies, 103*(1), 77-89.

Stouffer, D. B., & Bascompte, J. (2011). Compartmentalization increases food-web persistence. *Proceedings of the National Academy of Sciences, 108*(9), 3648-3652.

Suweis, S., Carr, J. A., Maritan, A., Rinaldo, A., & D'Odorico, P. (2015). Resilience and reactivity of global food security. *Proceedings of the National Academy of Sciences, 112*(22), 6902-6907.

Suweis, S., Konar, M., Dalin, C., Hanasaki, N., Rinaldo, A., & Rodriguez‐Iturbe, I. (2011). Structure and controls of the global virtual water trade network. *Geophysical Research Letters, 38*(10).

Suweis, S., Rinaldo, A., Maritan, A., & D'Odorico, P. (2013). Water-controlled wealth of nations. *Proceedings of the National Academy of Sciences, 110*(11), 4230-4233.

Tamea, S., Carr, J. A., Laio, F., & Ridolfi, L. (2014). Drivers of the virtual water trade. *Water Resources Research, 50*(1), 17-28.

Tu, C., Carr, J., & Suweis, S. (2016). A Data Driven Network Approach to Rank Countries Production Diversity and Food Specialization. *PLoS ONE, 11*(11), e0165941. https://www.ncbi.nlm.nih.gov/pubmed/27832118

Tu, C., Suweis, S., & D'Odorico, P. (2019). Impact of globalization on the resilience and sustainability of natural resources. *Nature Sustainability*.

Vågsholm, I., Arzoomand, N. S., & Boqvist, S. (2020). Food security, safety, and sustainability—getting the trade-offs right. *Frontiers in Sustainable Food Systems, 4*, 16.

Watts, D. J., & Strogatz, S. H. (1998). Collective dynamics of 'small-world' networks. *Nature, 393*(4), 440-442.

Xu, H., Niu, N., Li, D. M., & Wang, C. J. (2024). A Dynamic Evolutionary Analysis of the Vulnerability of Global Food Trade Networks. *Sustainability, 16*(10). <Go to ISI>://WOS:001233219700001

Xu, Z., Li, Y., Chau, S. N., Dietz, T., Li, C., Wan, L., et al. (2020). Impacts of international trade on global sustainable development. *Nature Sustainability, 3*(11), 964-971.

Zurek, M., Ingram, J., Bellamy, A. S., Goold, C., Lyon, C., Alexander, P., et al. (2022). Food System Resilience: Concepts, Issues, and Challenges. *annual review of environment and resources, 47*, 511-534.